\documentclass[preprint,eqsecnum,aps, showpacs]{revtex4}
\usepackage{graphicx}%
\usepackage{dcolumn}
\usepackage{amsmath}
\usepackage{bm}
\begin{document}

\title{Two neutrino positron double beta decay of $^{106}$Cd for 0$^{+}\rightarrow $%
0$^{+}$ transition}
%\date{}
\author{A. Shukla}
\email{ashukla@phy.iitkgp.ernet.in}
\affiliation{Department of Physics and Meteorology, IIT Kharagpur-721302, India.\\
}
\author{P. K. Raina}
%\email{pkraina@phy.iitkgp.ernet.in}
\affiliation{Department of Physics and Meteorology, IIT Kharagpur-721302, India.\\
}
\author{R. Chandra}
%\email{ramesh_luphy@yahoo.com}
\affiliation{Department of Physics, University of Lucknow, Lucknow-226007, India.\\
}
\author{P. K. Rath}
%\email{pkrath_55@hotmail.com}
\affiliation{Department of Physics, University of Lucknow, Lucknow-226007, India.\\
}
\author{J. G. Hirsch} 
\affiliation{Instituto de Ciencias Nucleares, Universidad Nacional Aut\'onoma de
M\'exico, A.P. 70-543 M\'exico 04510 D.F.\\ }
%\maketitle

\begin{abstract}
The two neutrino positron double beta decay of $^{106}$Cd for 0$%
^{+}\rightarrow $ 0$^{+}$ transition has been studied in the
Hartree-Fock-Bogoliubov model in conjunction with the summation method. In
the first step, the reliability of the intrinsic wave functions of $^{106}$%
Cd and $^{106}$Pd nuclei has been tested by comparing the theoretically
calculated results for yrast spectra, reduced $B(E2$:$0^{+}\rightarrow
2^{+}) $ transition probabilities, quadrupole moments $Q(2^{+})$ and
gyromagnetic factors $g(2^{+})$ with the available experimental data. In the
second step, the nuclear transition matrix element $M_{2\nu }$ and the
half-life $T_{1/2}^{2\nu }$ for 0$^{+}\rightarrow $ 0$^{+}$ transition have
been calculated with these wave functions. Moreover, we have studied the
effect of deformation on nuclear transition matrix element M$_{2\nu }.$
\end{abstract}
% PACS, the Physics and Astronomy Classification Scheme.

\maketitle 

\section{ Introduction}

The nuclear double beta ($\beta \beta $) decay, one of the rarest processes of
the nature, is characterized by two modes. They are the two neutrino double beta
(2$\nu $ $\beta \beta $) decay and the neutrinoless double beta (0$\nu $ $\beta
\beta $) decay. These modes can be classified into double electron ($\beta
^{-}\beta ^{-}$) emission, double positron $(\beta ^{+}\beta ^{+})$ emission,
electron-positron conversion $(\beta ^{+}EC)$ and double electron capture
$(ECEC)$. The later three processes are energetically competing and we shall
refer to them as positron double beta decay (e$^{+}$DBD) modes.  If  the 0$\nu
$ $\beta \beta $ decay is observed, the e$^{+}$DBD processes would play a
crucial role in discriminating the finer issues like dominance of Majorana
neutrino mass or the right handed current. The theoretical implications and
experimental aspects of e$^{+}$DBD modes have been widely reviewed over the
past years [1-8].

The half-lives of many $\beta ^{-}\beta ^{-}$ emitters are  shorter, compared
with the other modes, due to a larger available phase space.  For this reason
they were the natural choice for the experimental observation to start with.
However, the experimental sensitivity of $\beta ^{-}\beta ^{-}$ decay mode gets
limited because of the presence of electron background. On the other hand, from
the experimental point of view, the e$^{+}$DBD modes are relatively easier to
be separated from the background contaminations. Moreover, the e$^{+}$DBD modes
are also attractive due to the possibility to detect the coincidence signals
from four $\gamma $-rays, two $\gamma $-rays and one $\gamma $-ray for $ \beta
^{+}\beta ^{+}$, $\beta ^{+}EC$, $ECEC$ modes respectively. The {\it Q} -value
for the 2$\nu $ $ECEC$ mode can be large enough (up to 2.8 MeV) but the
detection of the 0$^{+}\rightarrow $0$^{+}$ transition is difficult since only
X-rays are emitted.

There have been very few experimental attempts for determining the half-lives
of 2$\nu $ e$^{+}$DBD modes even for the best candidate $^{106}$ Cd [9-16] but
one of the latest observations is very close to the predictions for $\beta
^{+}EC$ mode \cite{bel99}. With improved sensitivity in detection systems of
the planned bigger Osaka-OTO experiment \cite{ito97}, it is expected that
2$\nu $ e$^{+}$DBD modes will be in observable range in near future
\cite{zub01}. Hence, a timely reliable prediction of the half-life of
$^{106}$Cd decay will be helpful in designing of an experimental set up and
analysis of data.

Rosen and Primakoff were the first to study the 2$\nu $ e$^{+}$DBD modes
theoretically \cite{ros65}. Later on, Kim and Kubodera estimated the
half-lives of all the three modes with modified nuclear transition matrix
elements (NTMEs) and non-relativistic phase space factors \cite{kim83}. Abad 
{\it et al.} performed similar calculations using relativistic Coulomb wave
functions \cite{aba84}. In the mean time, the QRPA emerged as the most
successful model in explaining the quenching of NTMEs by incorporating the
particle-particle part of the effective nucleon-nucleon interaction in the
proton-neutron channel and the observed $T_{1/2}^{2\nu }$ of several 2$\nu $ 
$\beta \beta $ decay emitters were reproduced successfully \cite{suh98}.
Staudt {\it et al.} used QRPA model for evaluating 2$\nu $ $\beta ^{+}\beta
^{+}$ decay transition matrix elements \cite{staudt91}. Subsequently, the 2$%
\nu $ e$^{+}$DBD modes were studied in QRPA and its extensions [13, 22-26],
SU(4)$_{\sigma \tau }$ \cite{rum98} and SSDH \cite{civ98}.

A vast amount of data concerning the level energies as well as electromagnetic
properties have been compiled through experimental studies involving in-beam
$\gamma $-ray spectroscopy over the past years. Hence,  there is no need to
study the $\beta \beta $ decay  as an isolated nuclear process. The
availability of data permits a rigorous and detailed critique of the
ingredients of the microscopic framework that seeks to provide a description of
nuclear $\beta \beta $ decay. However, most of the calculations of e$^{+}$DBD
transition matrix elements performed so far but for the work of Barabash {\it
et al.}  \cite{bar96} and Suhonen {\it et al.} \cite{suh01} do not fully
satisfy this criterion.

The nuclear structure in the mass region {\it A}=100 offers a nice example of
shape transition i.e. sudden onset of deformation at neutron number {\it N
}=60. The nuclei are soft vibrators for  $N<$ 60 and quasi rotors for $N>$ 60. The
nuclei with neutron number {\it N} =60 are transitional nuclei. Hence, it is
expected that deformation degrees of freedom will play some crucial role in the
structure of $^{106}$Pd and $ ^{106}$Cd nuclei. Further, the pairing of like
nucleons plays an important role in all $\beta \beta $ decay emitters, which
are even-{\it Z} and even- {\it N} nuclei. Hence, it is desirable to have a
framework for the study of $ \beta \beta $ decay in which the pairing and
deformation degree of freedom are treated on equal footing in its formalism.
The Projected Hartree-Fock-Bogoliubov (PHFB) model is a very reasonable choice
which fulfills these requirements. The successful study of shape transition
vis-\'{a}-vis electromagnetic properties of various nuclei in PHFB model
[29-32] using pairing plus quadrupole-quadrupole (PPQQ) \cite{bara68}
interaction motivates us to apply the HFB wave functions to study the 2$\nu $
e$^{+}$DBD modes of $^{106}$Cd.

Further, it has been shown that there exists an inverse correlation between the
Gamow-Teller strength and the quadrupole moment \cite{Aue93,Tro96}. It is well
known that the pairing degree of freedom accounts for the preference of nuclei
to have a spherical form,  whereas the quadrupole-quadrupole ({\it QQ})
interaction increases the collectivity in the nuclear intrinsic wave functions
and makes the nucleus deformed. Hence, the PHFB model using the PPQQ
interaction is a convenient choice to examine the explicit role of deformation
on NTME $ M_{2\nu }.$

Our aim is to study the 2$\nu $ e$^{+}$DBD transition of $^{106}$Cd $%
\rightarrow $ $^{106}$Pd for 0$^{+}\rightarrow $0$^{+}$ transition together
with other observed nuclear properties using the PHFB model. In PHFB\ model,
the NTME $M_{2\nu }$ is usually calculated using the closure approximation. In
the present calculation, we have avoided the closure approximation by making
use of the summation method \cite{civ93}. In Sec. II, we briefly outline the
theoretical framework. In Sec. III, the reliability of the wave functions is
first established by calculating the yrast spectra, reduced
$B(E2:0^{+}\rightarrow 2^{+})$ transition probabilities, static quadrupole
moments $Q(2^{+})$ and $g$-factors $g(2^{+})$ of both parent $ ^{106}$Cd and
daughter $^{106}$Pd nuclei and by comparing them with the available
experimental data. The half-lives of 2$\nu $ e$^{+}$DBD modes for 0
$^{+}\rightarrow $0$^{+}$ transition have been given as prediction. The role of
deformation on NTME $M_{2\nu }$ has also been studied. We present the
conclusions in Sec. IV.

\section{Theoretical framework} 

The theoretical formalism to calculate the half-lives of 2$\nu $ e$^{+}$DBD
modes has been given by Doi {\it et al} \cite{doi92} and Suhonen {\it et al} \cite{suh98}.
Hence, we briefly outline steps of the above derivations for clarity in
notation following Doi {\it et al.} \cite{doi92}. Details of the mathematical
expressions used to calculate electromagnetic properties are given by Dixit
{\it et al.} \cite{bhanu}.

The half-life of the 2$\nu $ e$^{+}$DBD mode for the 0$^{+}\rightarrow $ 0$%
^{+}$ transition is given by

\begin{equation}
\left[ T_{1/2}^{2\nu }(0^{+}\rightarrow 0^{+})\right] ^{-1}=G_{2\nu }\left|
M_{2\nu }\right| ^{2}
\end{equation}
where the integrated kinematical factor $\ G_{2\nu }$\ can be calculated
with good accuracy \cite{doi92} and the NTME $M_{2\nu }$ is given by 
\begin{eqnarray}
M_{2\nu } &=&\sum\limits_{N}\frac{\langle 0_{F}^{+}||{\bm \sigma }\tau
^{-}||1_{N}^{+}\rangle \langle 1_{N}^{+}||{\bm \sigma }\tau
^{-}||0_{I}^{+}\rangle }{E_{N}-(E_{I}+E_{F})/2} \\
&=&\sum\limits_{N}\frac{\langle 0_{F}^{+}||{\bm \sigma }\tau
^{-}||1_{N}^{+}\rangle \langle 1_{N}^{+}||{\bm \sigma }\tau
^{-}||0_{I}^{+}\rangle }{E_{0}+E_{N}-E_{I}}
\end{eqnarray}
where $\ $%
\begin{eqnarray}
E_{0} &=&\frac{1}{2}\left( E_{I}-E_{F}\right)  \nonumber \\
&=&\frac{1}{2}Q_{\beta \beta }+m_{e}=\frac{1}{2}W_{0}
\end{eqnarray}
Here, $W_{0}$ is the total energy released and is given by 
\begin{eqnarray}
W_{0} &=&E_{I}-E_{F} \\
W_{0}(\beta ^{+}\beta ^{+}) &=&Q_{\beta ^{+}\beta ^{+}}+2m_{e} \\
W_{0}(\beta ^{+}EC) &=&Q_{\beta ^{+}EC}+e_{b} \\
W_{0}(ECEC) &=&Q_{ECEC}-2m_{e}+e_{b1}+e_{b2}
\end{eqnarray}

The summation over intermediate states can be completed using the summation
method \cite{civ93} and the $M_{2\nu }$ can be written as 
\begin{equation}
M_{2\nu }=\frac{1}{E_{0}}\left\langle 0_{F}^{+}\left| \sum_{m}(-1)^{m}\Gamma
_{-m}F_{m}\right| 0_{I}^{+}\right\rangle 
\end{equation}
where the Gamow-Teller (GT) operator $\Gamma _{m}$ is given by 
\begin{equation}
\Gamma _{m}=\sum_{s} {\bm \sigma_{ms} }\tau_{s} ^{-}
\end{equation}
and 
\begin{equation}
F_{m}=\sum_{\lambda =0}^{\infty }\frac{(-1)^{\lambda }}{E_{0}^{\lambda }}%
D_{\lambda }\Gamma _{m}
\end{equation}
with 
\begin{equation}
D_{\lambda }\Gamma _{m}=\left[ H,\left[ H,........,\right. \left[ H,\Gamma
_{m}\right] .......\right] ^{(\lambda \hbox{~times})}  \label{eqcom}
\end{equation}
In the present work, we use a Hamiltonian with PPQQ type \cite{bara68} of
effective two-body interaction. Explicitly, the Hamiltonian is written as 
\begin{equation}
{H}=H_{sp}+V(P)+\chi _{qq}V(QQ)
\end{equation}
where $H_{sp}$ denotes the single particle Hamiltonian. The pairing part of
the effective two-body interaction $V(P)$ is written as 
\begin{equation}
V{(}P{)}=-\left( \frac{G}{4}\right) \sum\limits_{\alpha \beta
}(-1)^{j_{\alpha }+j_{\beta }-m_{\alpha }-m_{\beta }}a_{\alpha }^{\dagger
}a_{\bar{\alpha}}^{\dagger }a_{\bar{\beta}}a_{\beta }
\end{equation}
where $\alpha $ denotes the quantum numbers ({\it nljm}). The state $\bar{%
\alpha}$ is same as $\alpha $ but with the sign of {\it m} reversed. The 
{\it QQ} part of the effective interaction $V(QQ)$\ is given by 
\begin{equation}
V(QQ)=-\left( \frac{\chi }{2}\right) \sum\limits_{\alpha \beta \gamma \delta
}\sum\limits_{\mu }(-1)^{\mu }\langle \alpha |q_{\mu }^{2}|\gamma \rangle
\langle \beta |q_{-\mu }^{2}|\delta \rangle \ a_{\alpha }^{\dagger }a_{\beta
}^{\dagger }\ a_{\delta }\ a_{\gamma }
\end{equation}
where 
\begin{equation}
{q_{\mu }^{2}}=\left( \frac{16\pi }{5}\right) ^{1/2}r^{2}Y_{\mu }^{2}(\theta
,\phi )
\end{equation}
The $\ \chi _{qq}$ is an arbitrary parameter and the final results are
obtained by setting the $\ \chi _{qq}$ = 1. The purpose of introducing $\chi
_{qq}$ is to study the role of deformation by varying the strength of {\it QQ%
} interaction.

When the GT operator commutes with the effective two-body interaction,
the Eq. (\ref{eqcom}) can be further simplified to 
\begin{equation}
M_{2\nu }=\sum\limits_{\pi ,\nu }\frac{\langle 0_{F}^{+}||{\bm \sigma}
.{\bm \sigma }\tau ^{-}\tau ^{-}||0_{I}^{+}\rangle }{E_{0}+\varepsilon (n_{\nu
},l_{\nu },j_{\nu })-\varepsilon (n_{\pi },l_{\pi },j_{\pi })}
\end{equation}

In the case of pseudo SU(3) model [38-40], the GT operator commutes with the
two-body interaction and the energy denominator is a well-defined quantity
without any free parameter. It has been evaluated exactly for 2$\nu $ $\beta
^{-}\beta ^{-}$ \cite{cas94,hir95} and 2$\nu $ $ECEC$ modes 
\cite{cer99} in the context of pseudo SU(3) scheme. However, 
in the present case, the
model Hamiltonian is not isospin symmetric. Hence, the energy denominator is
not as simple as in Eq. (2.17). But the violation of isospin symmetry for
the QQ part of our model Hamiltonian is negligible, as will be evident from
the parameters of the two-body interaction given later and the violation
of isospin symmetry for the pairing part of the two-body interaction is
presumably small. Under these assumptions, the expression to calculate the
NTME $M_{2\nu }$ of e$^{+}$DBD modes for 0$^{+}\rightarrow $ 0$^{+}$
transition in PHFB\ model is obtained as follows.

The essential idea behind the HFB theory is to transform particle
coordinates to quasiparticle coordinates through general Bogoliubov
transformation such that the quasiparticles are relatively weakly
interacting. Essentially, the Hamiltonian $H$ is expressed as

\begin{equation}
H=E_{0}+H_{qp}+H_{qp-int}
\end{equation}
where $E_{0}$ is the energy of the quasiparticle vacuum, $H_{qp}$ is the
elementary quasiparticle excitations and $H_{qp-int}$ is a weak interaction
between the quasiparticles. In HFB theory, the interaction between the
quasiparticles is usually neglected and the hamiltonian $H$ is approximated
by an independent quasiparticle hamiltonian. In time dependent HFB (TDHFB)
or the quasiparticle random phase approximation (QRPA), some effects of
quasiparticle interaction can be included. The axially symmetric intrinsic
HFB state with {\it K}=0 can be written as 
\begin{equation}
{|\Phi _{0}\rangle }=\prod\limits_{im}(u_{im}+v_{im}b_{im}^{\dagger }b_{i%
\bar{m}}^{\dagger })|0\rangle
\end{equation}
where the creation operators $\ b_{im}^{\dagger }$\ and $\ b_{i\bar{m}%
}^{\dagger }$\ are given by 
\begin{equation}
b{_{im}^{\dagger }}=\sum\limits_{\alpha }C_{i\alpha ,m}a_{\alpha m}^{\dagger
}\quad \hbox{and}{\rm \quad }b_{i\bar{m}}^{\dagger }=\sum\limits_{\alpha
}(-1)^{l+j-m}C_{i\alpha ,m}a_{\alpha ,-m}^{\dagger }
\end{equation}
Using the standard projection technique, a state with good angular momentum $%
{\bf J}$ is obtained from the HFB intrinsic state through the following
relation.

\begin{eqnarray}
{|\Psi _{MK}^{J}\rangle } &=&P_{MK}^{J}|\Phi _{K}\rangle  \nonumber \\
&=&\left[ \frac{(2J+1)}{{8\pi ^{2}}}\right] \int D_{MK}^{J}(\Omega )R(\Omega
)|\Phi _{K}\rangle d\Omega
\end{eqnarray}
where $\ R(\Omega )$\ and $\ D_{MK}^{J}(\Omega )$\ are the rotation operator
and the rotation matrix respectively.

Finally, one obtains the following expression for NTMEs of e$^{+}$DBD modes

\begin{eqnarray}
M_{2\nu } &=&\sum\limits_{\pi ,\nu }\frac{\langle {\Psi _{00}^{J_{f}=0}}||%
{\bm \sigma} .{\bm \sigma }\tau ^{-}\tau ^{-}||{\Psi _{00}^{J_{i}=0}}\rangle }{%
E_{0}+\varepsilon (n_{\nu },l_{\nu },j_{\nu })-\varepsilon (n_{\pi },l_{\pi
},j_{\pi })}  \nonumber \\
&=&[n_{Z-2,N+2}^{J_{f}=0}n_{Z,N}^{J_{i}=0}]^{-1/2}\int\limits_{0}^{\pi
}n_{(Z,N),(Z-2,N+2)}(\theta )  \nonumber \\
&&\times \sum\limits_{\alpha \beta \gamma \delta }\frac{\left\langle \alpha
\beta \left| {\bm \sigma }_{1}.{\bm \sigma }_{2}\tau ^{-}\tau ^{-}\right|
\gamma \delta \right\rangle }{E_{0}+\varepsilon _{\alpha }(n_{\nu },l_{\nu
},j_{\nu })-\varepsilon _{\gamma }(n_{\pi },l_{\pi },j_{\pi })}%
\sum_{\varepsilon \eta }\left[ \left( 1+F_{Z,N}^{(\nu )}(\theta
)f_{Z-2,N+2}^{(\nu )*}\right) \right] _{\varepsilon \alpha
}^{-1}(f_{Z-2,N+2}^{(\nu )*})_{\varepsilon \beta }  \nonumber \\
&&\times \left[ \left( 1+F_{Z,N}^{(\pi )}(\theta )f_{Z-2,N+2}^{(\pi
)*}\right) \right] _{\gamma \eta }^{-1}(F_{Z,N}^{(\pi )*})_{\eta \delta
}\sin \theta d\theta  \label{eq1}
\end{eqnarray}
where

\begin{equation}
n^{J}=\int\limits_{0}^{\pi }\{\det [1+F^{(\pi )}(\theta )f^{(\pi )\dagger
}]\}^{1/2}\times \{\det [1+F^{(\nu )}(\theta )f^{(\nu )\dagger
}]\}^{1/2}d_{00}^{J}(\theta )\sin (\theta )d\theta
\end{equation}
and

\begin{equation}
n_{(Z,N),(Z-2,N+2)}(\theta )=\{\det [1+F_{Z,N}^{(\pi )}(\theta
)f_{Z-2,N+2}^{(\pi )\dagger }]\}^{1/2}\times \{\det [1+F_{Z,N}^{(\nu
)}(\theta )f_{Z-2,N+2}^{(\nu )\dagger }]\}^{1/2}
\end{equation}
with

\begin{equation}
\lbrack F_{Z,N}(\theta )]_{\alpha \beta }=\sum_{m_{\alpha }^{^{\prime
}}m_{\beta }^{^{\prime }}}d_{m_{\alpha },m_{\alpha }^{^{\prime
}}}^{j_{\alpha }}(\theta )d_{m_{\beta },m_{\beta }^{^{\prime }}}^{j_{\beta
}}(\theta )f_{j_{\alpha }m_{\alpha }^{^{\prime }},j_{\beta }m_{\beta
}^{^{\prime }}}  \label{eq2}
\end{equation}
and

\begin{equation}
\lbrack f_{Z,N}]_{\alpha \beta }=\sum_{i}C_{ij_{\alpha },m_{\alpha
}}C_{ij_{\beta },m_{\beta }}\left( v_{im_{\alpha }}/u_{im_{\alpha }}\right)
\delta _{m_{\alpha },-m_{\beta }}
\end{equation}
Here $\pi (\nu )$ stands for the proton (neutron) of nuclei involved in 2$%
\nu $ e$^{+}$DBD. The results of PHFB calculations which are summarized by
the amplitudes $(u_{im},v_{im})$ and the expansion coefficients $C_{ij,m}$
are used to setup the matrices for $[F_{Z,N}(\theta )]_{\alpha \beta }$ and $%
[f_{Z,N}]_{\alpha \beta }$ given by Eqs. (2.25) and (2.26) respectively.
Finally, the required NTME $M_{2\nu }$ is calculated in a straight forward
manner using Eq. (\ref{eq1}) with 20 point gaussian quadrature points in the
range (0, $\pi $).

\section{Results and discussions}

The model space, single particle energies (SPE's) and two-body interactions
are same as our earlier calculation on 2$\nu $ $\beta \beta $ decay of $%
^{100}$Mo for 0$^{+}\rightarrow $0$^{+}$ transition \cite{bhanu}.
We include a brief discussion of them in the following for convenience.
We have treated the doubly even nucleus $^{76}$Sr ({\it N}={\it Z}=38) as an
inert core with the valence space spanned by the orbits 1{\it p}$_{1/2},$ 2%
{\it s}$_{1/2,}$ 1{\it d}$_{3/2}$, 1{\it d}$_{5/2}$, 0{\it g}$_{7/2}$, 0{\it %
g}$_{9/2}$ and 0{\it h}$_{11/2}$ for protons and neutrons. The orbit 1{\it p}%
$_{1/2}$ has been included in the valence space to examine the role of the 
{\it Z}=40 proton core vis-a-vis the onset of deformation in the highly
neutron rich isotopes.

The set of single particle energies (SPE's) used here are (in MeV) $%
\varepsilon $(1{\it p}$_{1/2}$)=-0.8, $\varepsilon $(0{\it g}$_{9/2}$)=0.0, $%
\varepsilon $(1{\it d}$_{5/2}$)=5.4, $\varepsilon $(2{\it s}$_{1/2}$)=6.4, $%
\varepsilon $(1{\it d}$_{3/2}$)=7.9, $\varepsilon $(0{\it g}$_{7/2}$)=8.4
and $\varepsilon $(0{\it h}$_{11/2}$)=8.6 for proton and neutrons. This set
of SPE's but for the $\varepsilon $(0{\it h}$_{11/2}$), which is slightly
lowered, has been employed in a number of successful shell model \cite
{verg71,fedm78} as well as variational model [29-32] calculations for
nuclear properties in the mass region {\it A}=100. The strengths of the
pairing interaction is fixed through the relation $G_{p}$ =30/{\it A} MeV
and $G_{n}$=20/{\it A} MeV, which are same as used by Heestand {\it et al.} 
\cite{hees69} to explain the experimental $g(2^{+})$ data of some even-even
Ge, Se, Mo, Ru, Pd, Cd and Te isotopes in Greiner's collective model \cite
{grei66}. The strengths of the like particle components of the {\it QQ}
interaction are taken as: $\chi _{pp}$ = $\chi _{nn}$ = 0.0105 MeV {\it b}$%
^{-4}$, where {\it b} is oscillator parameter. The strength of
proton-neutron ({\it pn}) component of the {\it QQ} interaction $\chi _{pn}$
is varied so as to reproduce the experimentally observed excitation energy
of the 2$^{+}$ state $E_{2^{+}}$ of $^{106}$Cd and $^{106}$Pd as closely as
possible. The $\chi _{pn}$ has been fixed to be 0.0151 and 0.0145 MeV {\it b}%
$^{-4}$ for $^{106}$Cd and $^{106}$Pd respectively. Thus for a given model
space, SPE's, $G_{p}$, $G_{n}$ and $\chi _{pp}$, we have fixed $\chi _{pn}$
through the experimentally available energy spectra. These values for the
strength of the {\it QQ} interaction are comparable to those suggested by
Arima on the basis of an empirical analysis of the effective two-body
interactions \cite{arim81}.

We have varied the $\chi _{pn}$ to obtain the yrast spectra of $^{106}$Cd
and $^{106}$Pd in optimum agreement with experimental results \cite{sakai}.
We have taken the theoretical spectra to be the optimum if the excitation
energy of the 2$^{+}$ state $E_{2^{+}}$ is reproduced as closely as possible
in comparison to the experimental results. Theoretically calculated intrinsic 
quadrupole moments $\left\langle
Q_{0}^{2}\right\rangle $ and yrast energies for the $E_{2^{+}}$ to $%
E_{6^{+}} $ levels of $^{106}$Cd and $^{106}$Pd for $\chi _{pn}$= 0.0142 to
0.0154 are presented in Table I. In the case of $^{106}$Cd, the $\left\langle
Q_{0}^{2}\right\rangle $ increases by 5.6267 units and the $E_{2^{+}}$
decreases by 0.1880 MeV as the $\chi _{pn}$ is varied from 0.0142 to 0.0154
MeV {\it b}$^{-4}$. For the same variation in $\chi _{pn},$ the $%
\left\langle Q_{0}^{2}\right\rangle $ increases by 3.7314 units and the $%
E_{2^{+}}$ decreases by 0.1109 MeV in the case of $^{106}$Pd. This observed
inverse correlation between $\left\langle Q_{0}^{2}\right\rangle $ and $%
E_{2^{+}}$, is understandable as there is an enhancement in the collectivity
of the intrinsic state with the increase of $\left| \chi _{pn}\right| ,$ the 
$E_{2^{+}}$ decreases$.$ This is known as Grodzins's rule \cite{grodzin}.
The theoretically calculated $E_{2^{+}}$ for $^{106}$Cd is 0.6220 MeV
corresponding to $\chi _{pn}$= 0.0151 MeV {\it b}$^{-4}$ in comparison to
the experimentally observed value 0.6327 MeV. In case of $^{106}$Pd, the
theoretically calculated $E_{2^{+}}$ for $\chi _{pn}$= 0.0145 MeV {\it b}$%
^{-4}$ is 0.5036 MeV in comparison to the observed value of 0.5119 MeV. All
these input parameters are kept fixed for calculation of spectroscopic
properties as well as the NTMEs discussed below.

The calculated as well as the experimentally observed values of the reduced $%
B(E2$:$0^{+}\rightarrow 2^{+})$ transition probabilities, static quadrupole
moments $Q(2^{+})$, and the gyromagnetic factors $g(2^{+})$ have been
presented in Table II . We have calculated $B(E2)$ values for effective
charges $e_{eff}$=0.40, 0.50, and 0.60, which are displayed in columns 2 to
4, respectively. The experimentally observed values are displayed in column
5. It is noticed that the calculated and the observed $B(E2)$ \cite{ram87}
values are in excellent agreement for $e_{eff}$=0.5. The theoretically
calculated $Q(2^{+})$ are tabulated in columns 6 to 8 for the same effective
charges as given above. The experimental $Q(2^{+})$ results \cite{rag89} are
given in column 9. It can be seen that for the same effective charge 0.5,
the calculated values are close to the experimental limit in case of $^{106}$%
Pd while the agreement between the calculated and experimental values is off
for $^{106}$Cd. The $g(2^{+})$ values are calculated with $g_{l}^{\pi }=$%
1.0, $g_{l}^{\nu }$=0.0, and $g_{s}^{\pi }=g_{s}^{\nu }=$0.60. The
calculated $g(2^{+})$ is 0.370 nm and 0.466 nm for $^{106}$Cd and $^{106}$Pd
respectively. The theoretically calculated and experimentally observed $%
g(2^{+})$ values are in good agreement for $^{106}$Cd and slightly off by
0.047 nm for $^{106}$Pd from the upper limit given by Raghavan \cite{rag89}.
The overall agreement between the calculated and observed electromagnetic
properties of $^{106}$Cd and $^{106}$Pd suggests that the PHFB wave
functions generated by fixing $\chi _{pn}$ to reproduce the yrast spectra
are quite reliable.

The 2$\nu $ e$^{+}$DBD of $^{106}$Cd for the 0$^{+}\rightarrow $ 0$^{+}$
transition has been investigated by very few experimental groups, whereas
some theoretical investigations have been made using the QRPA and its 
extensions [13,21-26], SU(4)$_{\sigma \tau }$ \cite{rum98} and SSDH \cite{civ98}. In
Table III, we have compiled all the available experimental [9-16] and
theoretical results [13,21-28] along with our calculated $M_{2\nu }$ and
corresponding half-life $T_{1/2}^{2\nu }$. We have used phase space factors $%
G_{2\nu }=$4.991$\times $10$^{-26}$ yr$^{-1}$, 1.970$\times $10$^{-21}$yr$%
^{-1}$ and 1.573$\times $10$^{-20}$ yr$^{-1}$ for 2$\nu $ $\beta ^{+}\beta
^{+}$, 2$\nu $ $\beta ^{+}EC$ and 2$\nu $ $ECEC$ modes respectively as given
by Doi {\it et al.} \cite{doi92}. The phase space integral has been
evaluated for $\ g_{A}$= 1.261 by Doi {\it et al.} \cite{doi92}. However, in
heavy nuclei it is more justified to use the nuclear matter value of $\ g_{A}
$ around 1.0. Hence, the theoretical $T_{1/2}^{2\nu }$ are calculated for $\
g_{A}$=1.0 and 1.261. We have presented only the theoretical $T_{1/2}^{2\nu }
$ for those models for which no direct or indirect information about $%
M_{2\nu }$ or $G_{2\nu }$ is available to us.

In column 3 of Table III, we have presented the experimentally observed
limits on half-lives $T_{1/2}^{2\nu }$. In comparison to the theoretically
predicted $T_{1/2}^{2\nu },$ the present experimental limits for 0$%
^{+}\rightarrow $0$^{+}$ transition of $^{106}$Cd are smaller by a factor of
10$^{5-7}$ in case of 2$\nu $ $\beta ^{+}\beta ^{+}$ mode but are quite
close for 2$\nu $ $\beta ^{+}EC$ and 2$\nu $ $ECEC$ modes. The half-life $%
T_{1/2}^{2\nu }$ calculated in PHFB model using the summation method differs
from all the existing calculations. The presently calculated NTME $M_{2\nu }$
is smaller than the recently given results in QRPA(WS) model of Suhonen and
Civitarese \cite{suh01} by a factor of 2 approximately for all the three
modes. The theoretical $M_{2\nu }$ values of PHFB model and SU(4)$_{\sigma
\tau }$ \cite{rum98} again differ by a factor of 2 approximately for the 2$%
\nu $ $\beta ^{+}EC$ and 2$\nu $ $ECEC$ modes. On the other hand, the $%
M_{2\nu }$ calculated in our PHFB model is smaller than the values of Hirsch 
{\it et al.} \cite{hir94} by a factor of 3 approximately in case of 2$\nu $ $%
\beta ^{+}\beta ^{+}$ and 2$\nu $ $ECEC$ modes while for 2$\nu $ $\beta
^{+}EC$ mode the results differ by a factor of 4 approximately. All the rest
of the calculations predict NTMEs, which are larger than our predicted $%
M_{2\nu }$ approximately by a factor of 7 \cite{suh93,toi97} to 10 \cite
{bar96}.

We have studied the role of deformation on $\left\langle
Q_{0}^{2}\right\rangle $ and $M_{2\nu }$ vis-a-vis the variation of the
strength of {\it pn} part of the {\it QQ} interaction $\chi _{qq}$. The
results are tabulated in Table IV. The $\left\langle Q_{0}^{2}\right\rangle $
of $^{106}$Cd and $^{106}$Pd remain almost constant as the $\chi _{qq}$ is
varied from 0.0 to 0.80. The $M_{2\nu }$ also remains almost constant as the 
$\chi _{qq}$ is changed from 0.0 to 0.80. As the $\chi _{qq}$ is further
changed from 0.80 to 1.20, the $\left\langle Q_{0}^{2}\right\rangle $
increases while the $M_{2\nu }$ decreases to 0.0417 having a fluctuation at
1.05. To quantify the effect of deformation on $M_{2\nu }$, we define a
quantity $D_{2\nu }$ as the ratio of $M_{2\nu }$ at zero deformation ($\chi
_{qq}=0$) and full deformation ($\chi _{qq}=1$). The $D_{2\nu }$ is given by 
\begin{equation}
D_{2\nu }=\frac{M_{2\nu }(\chi _{qq}=0)}{M_{2\nu }(\chi _{qq}=1)}
\end{equation}
The value of $D_{2\nu }$ is 2.09, which suggest that the $M_{2\nu }$ is
quenched by a factor of approximately 2 due to deformation effects.

It is evident from the above discussions that it is difficult to establish
the validity of different nuclear models presently employed to study 2$\nu $
e$^{+}$DBD due to limiting values in experimental results as well as
uncertainty in $g_{A}$. Further work is necessary both in the experimental
as well as the theoretical front to judge the relative applicability,
success and failure of various models used so far for the study of 2$\nu $ e$%
^{+}$DBD processes before they can have better predictive power for the 0$%
\nu $ e$^{+}$DBD modes.

\section{Conclusions}

We have tested the quality of HFB wave functions by comparing the
theoretically calculated results for a number of spectroscopic properties
namely yrast spectra, reduced $B(E2$:$0^{+}\rightarrow 2^{+})$ transition
probabilities, quadrupole moments $Q(2^{+})$ and $g$-factors $g(2^{+})$ of $%
^{106}$Cd and $^{106}$Pd with the available experimental data. The same HFB
wave functions are employed to calculate the NTME $M_{2\nu }$ and the
half-life $T_{1/2}^{2\nu }$ of $^{106}$Cd for 2$\nu $ $\beta ^{+}\beta ^{+}$%
, 2$\nu $ $\beta ^{+}EC$ and 2$\nu $ $ECEC$ modes. The values of $%
T_{1/2}^{2\nu }$ calculated in the PHFB model with the summation method are
larger than the previous calculations. The presently calculated NTME $%
M_{2\nu }$ is smaller than the recently given results in QRPA(WS) model of
Suhonen and Civitarese \cite{suh01} by a factor of 2 approximately for all
the three modes. The proton-neutron part of the PPQQ interaction that
reflects the deformations of intrinsic ground state, plays an important role
in the quenching of $M_{2\nu }$ by a factor of 2 approximately in this
particular case. The calculated 2$\nu $ $e^{+}$DBD decay half-lives are very
close to the experimentally observable limits for 2$\nu $ $\beta ^{+}EC$ and
2$\nu $ $ECEC$ modes. It is hoped that the calculated $T_{1/2}^{2\nu }$,
which is of the order of 10$^{21-23}$ yrs can be reached experimentally for 2%
$\nu $ $\beta ^{+}EC$ mode in near future \cite{bel99}.

{\bf Acknowledgments:} P. K. Rath would like to acknowledge the financial
support provided by CTS, Indian Institute of Technology, Kharagpur, India,
where part of this work was carried out. Further, RC is grateful to
CSIR, India for providing SRF vide award no. 9/107(222)/2KI/EMR-I.

\begin{table}
\caption{Variation in intrinsic quadrupole moment $\left\langle
Q_{0}^{2}\right\rangle $ and excitation energies (in MeV) of $J^{\pi }=$2$%
^{+}$, 4$^{+}$, and 6$^{+}$ yrast states of $^{106}$Cd and $^{106}$Pd nuclei
with change in $\chi _{pn}$, keeping fixed $G_{p}$=30/{\it A} MeV, $G_{n}$%
=20/{\it A} MeV, $\chi _{pp}=\chi _{nn}=$0.0105 MeV {\it b}$^{-4}$ and $%
\varepsilon ($0{\it h}$_{11/2})=$ 8.6 MeV.}

\begin{ruledtabular}
\begin{tabular}{cccccccc}

Nucleus & \multicolumn{6}{c}{Theo.} & Expt.\cite{sakai} \\ 

& $\chi _{pn}=$ & 0.0142 & 0.0145 & 0.0148 & 0.0151 & 0.0154 &  \\ 

\hline

$^{106}$Cd & $\left\langle Q_{0}^{2}\right\rangle $ & 43.3772 & 44.7355 & 
46.0289 & 47.3807 & 49.0039 &  \\ 

& $E_{2^{+}}$ & 0.7749 & 0.7339 & 0.6797 & {\bf 0.6220} & 0.5869 & {\bf %
0.6327} \\ 

& $E_{4^{+}}$ & 1.9024 & 1.8728 & 1.8022 & 1.7129 & 1.6690 & 1.4939 \\ 

& $E_{6^{+}}$ & 3.2993 & 3.3029 & 3.2389 & 3.1411 & 3.1089 & 2.4918 \\ 

\hline

$^{106}$Pd & $\left\langle Q_{0}^{2}\right\rangle $ & 51.4360 & 52.4295 & 
53.4325 & 54.2709 & 55.1674 &  \\ 

& $E_{2^{+}}$ & 0.5524 & {\bf 0.5036} & 0.4819 & 0.4500 & 0.4415 & {\bf %
0.5119} \\ 

& $E_{4^{+}}$ & 1.5706 & 1.4668 & 1.4269 & 1.3554 & 1.3435 & 1.2292 \\ 

& $E_{6^{+}}$ & 2.8526 & 2.7089 & 2.6655 & 2.5652 & 2.5620 & 2.0766 \\ 

\end{tabular}
\end{ruledtabular}
\end{table}

\begin{table}
\caption{Comparison of the calculated and experimentally observed
reduced transition probabilities $B(E2$:$0^{+}\rightarrow 2^{+}),$ static
quadrupole moments $Q(2^{+})$ and $g$-factors $g(2^{+})$. Here $B(E2)$ and $%
Q(2^{+})$ are calculated in units of $e^{2}$ b$^{2}$ and $e$ b, respectively
for effective charge $e_{p}=$1+$e_{eff}$ and $e_{n}=e_{eff}$. The $g(2^{+})$
has been calculated in units of nuclear magneton for $g_{l}^{\pi }$=1.0, $%
g_{l}^{\nu }$=0.0 and $g_{s}^{\pi }$=$g_{s}^{\nu }$=0.60. Corresponding
references for experimentally observed values are given in parentheses.}

\begin{ruledtabular}
\begin{tabular}{ccccccccccc}

Nucleus & \multicolumn{4}{c}{$B(E2:0^{+}\rightarrow 2^{+})$} & 
\multicolumn{4}{c}{$Q(2^{+})$} & \multicolumn{2}{c}{$g(2^{+})$} \\ 

& \multicolumn{3}{c}{Theo.} & Expt.\cite{ram87} & \multicolumn{3}{c}{Theo.}
& Expt. \cite{rag89} & Theo. & Expt. \cite{rag89} \\ 

& \multicolumn{3}{c}{$e_{eff}$} &  & \multicolumn{3}{c}{$e_{eff}$} &  &  &\\ 

& 0.40 & 0.50 & 0.60 &  & 0.40 & 0.50 & 0.60 &  &  &  \\ 

\hline

$^{106}$Cd & 0.334 & 0.426 & 0.531 & 0.410$\pm 0.$02$0$ & -0.52 & -0.59 & 
-0.66 & -0.28$\pm 0.08$ & 0.370 & 0.40$\pm 0.10$ \\ 

&  &  &  & 0.386$\pm 0.$05 &  &  &  &  &  &  \\ 

\hline

$^{106}$Pd & 0.407 & 0.520 & 0.647 & 0.610$\pm 0.090$ & -0.58 & -0.65 & -0.73
& -0.56$\pm 0.08$ & 0.466 & 0.398$\pm 0.021$ \\ 

&  &  &  & 0.656$\pm 0.035$ &  &  &  & -0.51$\pm 0.08$ &  & 0.30$\pm 0.06$

\end{tabular}
\end{ruledtabular}
\end{table}

\pagebreak 

\begin{table}
\caption{ Experimental limits on half-lives $T_{1/2}^{2\nu }($0$%
^{+}\rightarrow $0$^{+})$, theoretically calculated $M_{2\nu }$ and
corresponding $T_{1/2}^{2\nu }($0$^{+}\rightarrow $0$^{+})$ for 2$\nu $ $%
\beta ^{+}\beta ^{+},$ 2$\nu $ $\beta ^{+}EC$ and 2$\nu $ $ECEC$ modes of $%
^{106}$Cd. The numbers corresponding to (a) and (b) are calculated for $%
g_{A} $=1.261 and 1.0 respectively.}

\begin{ruledtabular}
\begin{tabular}{cccccccc}

Decay & \multicolumn{2}{c}{Experiment} & \multicolumn{5}{c}{Theory} \\ 

Mode & Ref. & $T_{1/2}^{2\nu }$ ( yr) & Ref. & Models & $\left| M_{2\nu
}\right| $ &  & $T_{1/2}^{2\nu }$ ( yr) \\ 

\hline

\multicolumn{1}{l}{$\beta ^{+}\beta ^{+}$} & \multicolumn{1}{l}{\cite{dan03}}
& \multicolumn{1}{l}{$>5.0 \times 10^{18}$} & \multicolumn{1}{l}{
Present} & \multicolumn{1}{l}{PHFB} & \multicolumn{1}{l}{0.081} & 
\multicolumn{1}{l}{a)} & \multicolumn{1}{l}{{\bf 307.58}$\times ${\bf 10}$%
^{25}$} \\ 

\multicolumn{1}{l}{} & \multicolumn{1}{l}{\cite{bel99}} & \multicolumn{1}{l}{%
$> 2.4 \times 10^{20**}$} & \multicolumn{1}{l}{} & 
\multicolumn{1}{l}{} & \multicolumn{1}{l}{} & \multicolumn{1}{l}{b)} & 
\multicolumn{1}{l}{{\bf 777.71}$\times ${\bf 10}$^{25}$} \\

& \cite{bar96} & $> 1.0 \times $10$^{19*}$ & \multicolumn{1}{l}{%
\cite{sto03}} & \multicolumn{1}{l}{SQRPA(l.b.)} & \multicolumn{1}{l}{0.61} & 
a) & \multicolumn{1}{l}{5.38$\times $10$^{25}$} \\ 

& \cite{dan96} & \multicolumn{1}{l}{$> 9.2\times 10^{17}$ }&  & 
\multicolumn{1}{l}{} & \multicolumn{1}{l}{} & b) & \multicolumn{1}{l}{13.60$%
\times $10$^{25}$} \\ 

& \cite{mit88} & \multicolumn{1}{l}{$> 5.0 \times 10^{17}$ }&  & 
\multicolumn{1}{l}{SQRPA(s.b.)} & \multicolumn{1}{l}{0.57} & a) & 
\multicolumn{1}{l}{6.16$\times $10$^{25}$} \\ 

& \cite{nor84} & \multicolumn{1}{l}{$>2.6 \times 10^{17*}$} &  & 
\multicolumn{1}{l}{} & \multicolumn{1}{l}{} & b) & \multicolumn{1}{l}{15.58$%
\times $10$^{25}$} \\ 

\multicolumn{1}{l}{} & \multicolumn{1}{l}{} & \multicolumn{1}{l}{} & 
\multicolumn{1}{l}{\cite{suh01}} & \multicolumn{1}{l}{QRPA(WS)} & 
\multicolumn{1}{l}{0.166} & \multicolumn{1}{l}{a)} & \multicolumn{1}{l}{72.71%
$\times $10$^{25}$} \\ 

\multicolumn{1}{l}{} & \multicolumn{1}{l}{} & \multicolumn{1}{l}{} & 
\multicolumn{1}{l}{} & \multicolumn{1}{l}{} & \multicolumn{1}{l}{} & 
\multicolumn{1}{l}{b)} & \multicolumn{1}{l}{183.84$\times $10$^{25}$} \\ 

\multicolumn{1}{l}{} & \multicolumn{1}{l}{} & \multicolumn{1}{l}{} & 
\multicolumn{1}{l}{} & \multicolumn{1}{l}{QRPA(AWS)} & \multicolumn{1}{l}{
0.722} & \multicolumn{1}{l}{a)} & \multicolumn{1}{l}{3.84$\times $10$^{25}$}\\ 

\multicolumn{1}{l}{} & \multicolumn{1}{l}{} & \multicolumn{1}{l}{} & 
\multicolumn{1}{l}{} & \multicolumn{1}{l}{} & \multicolumn{1}{l}{} & 
\multicolumn{1}{l}{b)} & \multicolumn{1}{l}{9.72$\times $10$^{25}$} \\ 

\multicolumn{1}{l}{} & \multicolumn{1}{l}{} & \multicolumn{1}{l}{} & 
\multicolumn{1}{l}{\cite{bar96}} & \multicolumn{1}{l}{QRPA(WS)} & 
\multicolumn{1}{l}{0.840} & \multicolumn{1}{l}{a)} & \multicolumn{1}{l}{2.84$%
\times $10$^{25}$} \\ 

&  &  &  & \multicolumn{1}{l}{} & \multicolumn{1}{l}{} & \multicolumn{1}{l}{
b)} & \multicolumn{1}{l}{7.18$\times $10$^{25}$} \\ 

&  &  &  & \multicolumn{1}{l}{QRPA(AWS)} & \multicolumn{1}{l}{0.780} & 
\multicolumn{1}{l}{a)} & \multicolumn{1}{l}{3.29$\times $10$^{25}$} \\ 

&  &  &  & \multicolumn{1}{l}{} & \multicolumn{1}{l}{} & \multicolumn{1}{l}{
b)} & \multicolumn{1}{l}{8.33$\times $10$^{25}$} \\ 

\multicolumn{1}{l}{} & \multicolumn{1}{l}{} & \multicolumn{1}{l}{} & 
\multicolumn{1}{l}{\cite{hir94}} & \multicolumn{1}{l}{QRPA} & 
\multicolumn{1}{l}{0.218} & \multicolumn{1}{l}{a)} & \multicolumn{1}{l}{42.2$%
\times $10$^{25}$} \\ 

&  &  &  &  &  & b) & \multicolumn{1}{l}{106.6$\times $10$^{25}$} \\ 

\multicolumn{1}{l}{} & \multicolumn{1}{l}{} & \multicolumn{1}{l}{} & 
\multicolumn{1}{l}{\cite{staudt91}} & \multicolumn{1}{l}{QRPA} & 
\multicolumn{1}{l}{} & \multicolumn{1}{l}{} & \multicolumn{1}{l}{4.94$\times 
$10$^{25}$} \\ 

\end{tabular}
\end{ruledtabular}
\end{table}

\addtocounter{table}{-1}%
\begin{table}
\caption{.....Continued}
%\begin{ruledtabular}
\begin{tabular}{llllllll}
\hline
\multicolumn{1}{l}{$\beta ^{+}EC$} & \multicolumn{1}{l}{\cite{dan03}} & 
\multicolumn{1}{l}{$>1.2 \times 10^{18}$ }& \multicolumn{1}{l}{
Present} & \multicolumn{1}{l}{PHFB} & \multicolumn{1}{l}{0.081} & 
\multicolumn{1}{l}{a)} & \multicolumn{1}{l}{{\bf 77.925}$\times ${\bf 10}$%
^{21}$} \\ 

\multicolumn{1}{l}{} & \multicolumn{1}{l}{\cite{bel99}} & \multicolumn{1}{l}{%
$> 4.1 \times $10$^{20}$} & \multicolumn{1}{l}{} & 
\multicolumn{1}{l}{} & \multicolumn{1}{l}{} & \multicolumn{1}{l}{b)} & 
\multicolumn{1}{l}{{\bf 197.03}$\times ${\bf 10}$^{21}$} \\ 

& \cite{bar96} & $> 0.66 \times $10$^{19*}$ & \multicolumn{1}{l}{%
\cite{sto03}} & \multicolumn{1}{l}{SQRPA(l.b.)} & \multicolumn{1}{l}{0.61} & 
a) & \multicolumn{1}{l}{1.36$\times $10$^{21}$} \\ 

& \cite{dan96} & $>2.6 \times 10^{17}$ &  & \multicolumn{1}{l}{}
& \multicolumn{1}{l}{} & b) & \multicolumn{1}{l}{3.44$\times $10$^{21}$} \\ 

& \cite{nor84} & $>5.7 \times 10^{17*}$ &  & \multicolumn{1}{l}{
SQRPA(s.b.)} & \multicolumn{1}{l}{0.57} & a) & \multicolumn{1}{l}{1.56$%
\times $10$^{21}$} \\ 

&  &  &  & \multicolumn{1}{l}{} & \multicolumn{1}{l}{} & b) & 
\multicolumn{1}{l}{3.94$\times $10$^{21}$} \\ 

\multicolumn{1}{l}{} & \multicolumn{1}{l}{} & \multicolumn{1}{l}{} & 
\multicolumn{1}{l}{\cite{suh01}} & \multicolumn{1}{l}{QRPA(WS)} & 
\multicolumn{1}{l}{0.168} & \multicolumn{1}{l}{a)} & \multicolumn{1}{l}{17.99%
$\times $10$^{21}$} \\ 

\multicolumn{1}{l}{} & \multicolumn{1}{l}{} & \multicolumn{1}{l}{} & 
\multicolumn{1}{l}{} & \multicolumn{1}{l}{} & \multicolumn{1}{l}{} & 
\multicolumn{1}{l}{b)} & \multicolumn{1}{l}{45.48$\times $10$^{21}$} \\ 

\multicolumn{1}{l}{} & \multicolumn{1}{l}{} & \multicolumn{1}{l}{} & 
\multicolumn{1}{l}{} & \multicolumn{1}{l}{QRPA(AWS)} & \multicolumn{1}{l}{
0.718} & \multicolumn{1}{l}{a)} & \multicolumn{1}{l}{0.98$\times $10$^{21}$}\\ 

\multicolumn{1}{l}{} & \multicolumn{1}{l}{} & \multicolumn{1}{l}{} & 
\multicolumn{1}{l}{} & \multicolumn{1}{l}{} & \multicolumn{1}{l}{} & 
\multicolumn{1}{l}{b)} & \multicolumn{1}{l}{2.49$\times $10$^{21}$} \\ 

\multicolumn{1}{l}{} & \multicolumn{1}{l}{} & \multicolumn{1}{l}{} & 
\multicolumn{1}{l}{\cite{rum98}} & \multicolumn{1}{l}{SU(4)$_{\sigma \tau }$}
& \multicolumn{1}{l}{0.1947} & \multicolumn{1}{l}{a)} & \multicolumn{1}{l}{
13.39$\times $10$^{21}$} \\ 

\multicolumn{1}{l}{} & \multicolumn{1}{l}{} & \multicolumn{1}{l}{} & 
\multicolumn{1}{l}{} & \multicolumn{1}{l}{} & \multicolumn{1}{l}{} & 
\multicolumn{1}{l}{b)} & \multicolumn{1}{l}{33.86$\times $10$^{21}$} \\ 

\multicolumn{1}{l}{} & \multicolumn{1}{l}{} & \multicolumn{1}{l}{} & 
\multicolumn{1}{l}{\cite{toi97}} & \multicolumn{1}{l}{RQRPA(WS)} & 
\multicolumn{1}{l}{0.550} & \multicolumn{1}{l}{a)} & \multicolumn{1}{l}{1.68$%
\times $10$^{21}$} \\ 

\multicolumn{1}{l}{} & \multicolumn{1}{l}{} & \multicolumn{1}{l}{} & 
\multicolumn{1}{l}{} & \multicolumn{1}{l}{} & \multicolumn{1}{l}{} & 
\multicolumn{1}{l}{b)} & \multicolumn{1}{l}{4.24$\times $10$^{21}$} \\ 

\multicolumn{1}{l}{} & \multicolumn{1}{l}{} & \multicolumn{1}{l}{} & 
\multicolumn{1}{l}{\cite{toi97}} & \multicolumn{1}{l}{RQRPA(AWS)} & 
\multicolumn{1}{l}{0.560} & \multicolumn{1}{l}{a)} & \multicolumn{1}{l}{1.62$%
\times $10$^{21}$} \\ 

\multicolumn{1}{l}{} & \multicolumn{1}{l}{} & \multicolumn{1}{l}{} & 
\multicolumn{1}{l}{} & \multicolumn{1}{l}{} & \multicolumn{1}{l}{} & 
\multicolumn{1}{l}{b)} & \multicolumn{1}{l}{4.09$\times $10$^{21}$} \\ 

\multicolumn{1}{l}{} & \multicolumn{1}{l}{} & \multicolumn{1}{l}{} & 
\multicolumn{1}{l}{\cite{bar96}} & \multicolumn{1}{l}{QRPA(WS)} & 
\multicolumn{1}{l}{0.840} & \multicolumn{1}{l}{a)} & \multicolumn{1}{l}{0.72$%
\times $10$^{21}$} \\ 

&  &  &  & \multicolumn{1}{l}{} & \multicolumn{1}{l}{} & \multicolumn{1}{l}{
b)} & \multicolumn{1}{l}{1.82$\times $10$^{21}$} \\ 

&  &  &  & \multicolumn{1}{l}{QRPA(AWS)} & \multicolumn{1}{l}{0.780} & 
\multicolumn{1}{l}{a)} & \multicolumn{1}{l}{0.83$\times $10$^{21}$} \\ 

&  &  &  & \multicolumn{1}{l}{} & \multicolumn{1}{l}{} & \multicolumn{1}{l}{
b)} & \multicolumn{1}{l}{2.11$\times $10$^{21}$} \\ 

\multicolumn{1}{l}{} & \multicolumn{1}{l}{} & \multicolumn{1}{l}{} & 
\multicolumn{1}{l}{\cite{hir94}} & \multicolumn{1}{l}{QRPA} & 
\multicolumn{1}{l}{0.352} & \multicolumn{1}{l}{a)} & \multicolumn{1}{l}{4.1$%
\times $10$^{21}$} \\ 

\multicolumn{1}{l}{} & \multicolumn{1}{l}{} & \multicolumn{1}{l}{} & 
\multicolumn{1}{l}{} & \multicolumn{1}{l}{} & \multicolumn{1}{l}{} & 
\multicolumn{1}{l}{b)} & \multicolumn{1}{l}{10.4$\times $10$^{21}$} \\ 

\multicolumn{1}{l}{} & \multicolumn{1}{l}{} & \multicolumn{1}{l}{} & 
\multicolumn{1}{l}{\cite{suh93}} & \multicolumn{1}{l}{QRPA(WS)} & 
\multicolumn{1}{l}{0.493-0.660} & \multicolumn{1}{l}{a)} & 
\multicolumn{1}{l}{(2.09-1.16)$\times $10$^{21}$} \\ 

\multicolumn{1}{l}{} & \multicolumn{1}{l}{} & \multicolumn{1}{l}{} & 
\multicolumn{1}{l}{} & \multicolumn{1}{l}{} & \multicolumn{1}{l}{} & 
\multicolumn{1}{l}{b)} & \multicolumn{1}{l}{(5.28-2.95)$\times $10$^{21}$}\\ 

\hline
\end{tabular}
%\end{ruledtabular}
\end{table}

\addtocounter{table}{-1}%
\begin{table}
\caption{.....Continued}
%\begin{ruledtabular}
\begin{tabular}{llllllll}
\hline
$ECEC$ & \cite{dan03} & $ > 5.8 \times 10^{17}$ & Present & PHFB & 
0.081 & a) & {\bf 97.593}$\times ${\bf 10}$^{20}$ \\ 

& \cite{zub03} & $> 1.0 \times $10$^{18}$ &  &  &  & b) & {\bf %
246.76}$\times ${\bf 10}$^{20}$ \\ 

& \cite{geo95a} & $> 5.8 \times $10$^{17}$ & \cite{sto03} & 
SQRPA(l.b.) & 0.61 & a) & 2.6$\times $10$^{20}$ \\ 

&  &  &  &  &  & b) & 6.57$\times $10$^{20}$ \\ 

&  &  &  & SQRPA(s.b.) & 0.57 & a) & 1.96$\times $10$^{20}$ \\ 

&  &  &  &  &  & b) & 4.96$\times $10$^{20}$ \\ 

&  &  & \cite{suh01} & QRPA(WS) & 0.168 & a) & 22.52$\times $10$^{20}$ \\ 

&  &  &  &  &  & b) & 56.95$\times $10$^{20}$ \\ 

&  &  &  & QRPA(AWS) & 0.718 & a) & 1.23$\times $10$^{20}$ \\ 

&  &  &  &  &  & b) & 3.12$\times $10$^{20}$ \\ 

&  &  & \cite{civ98} & SSDH(Theo) & 0.280 & a) & 8.11$\times $10$^{20}$ \\ 

&  &  &  &  &  & b) & 20.50$\times $10$^{20}$ \\ 

&  &  &  & SSDH(Exp) & 0.170 & a) & 22.00$\times $10$^{20}$ \\ 

&  &  &  &  &  & b) & 55.62$\times $10$^{20}$ \\ 

&  &  & \cite{rum98} & SU(4)$_{\sigma \tau }$ & 0.1947 & a) & 16.77$\times $%
10$^{20}$ \\ 

&  &  &  &  &  & b) & 42.40$\times $10$^{20}$ \\ 

&  &  & \cite{toi97} & RQRPA(WS) & 0.550 & a) & 2.10$\times $10$^{20}$ \\ 

&  &  &  &  &  & b) & 5.31$\times $10$^{20}$ \\ 

&  &  & \cite{toi97} & RQRPA(AWS) & 0.560 & a) & 2.03$\times $10$^{20}$ \\ 

&  &  &  &  &  & b) & 5.13$\times $10$^{20}$ \\ 

&  &  & \cite{bar96} & QRPA(WS) & 0.840 & a) & 0.90$\times $10$^{20}$ \\ 

\multicolumn{1}{c}{} & \multicolumn{1}{c}{} & \multicolumn{1}{c}{} &  &  & 
& b) & 2.28$\times $10$^{20}$ \\ 

\multicolumn{1}{c}{} & \multicolumn{1}{c}{} & \multicolumn{1}{c}{} &  & 
QRPA(AWS) & 0.780 & a) & 1.05$\times $10$^{20}$ \\ 

\multicolumn{1}{c}{} & \multicolumn{1}{c}{} & \multicolumn{1}{c}{} &  &  & 
& b) & 2.64$\times $10$^{20}$ \\ 

&  &  & \cite{hir94} & QRPA & 0.270 & a) & 8.7$\times $10$^{20}$ \\ 
&  &  &  &  &  & b) & 22.1$\times $10$^{20}$ \\ 

&  &  & \cite{suh93} & QRPA(WS) & 0.493-0.660 & a) & (2.62-1 .46)$\times $10$%
^{20}$ \\ 

&  &  &  &  &  & b) & (6.61-3.69)$\times $10$^{20}$ \\

\hline\hline
\end{tabular}
%\end{ruledtabular}
\footnotetext{* and ** denote half-life limit for 0$\nu $ + 2$\nu $ and 0$\nu $ + 2$\nu
+0\nu $M modes respectively.}
\end{table}

\pagebreak

\begin{table}
\caption{Effect of the variation in $\chi _{qq}$ on $\left\langle
Q_{0}^{2}\right\rangle $ and $M_{2\nu }.$}

\begin{ruledtabular}
\begin{tabular}{cccccccc}

$\chi _{qq}$ & \multicolumn{3}{c}{$^{106}$Cd} & \multicolumn{3}{c}{$^{106}$Pd
} & $\left| M_{2\nu }\right| $ \\ 

& $\left\langle Q_{0}^{2}\right\rangle _{\pi }$ & $\left\langle
Q_{0}^{2}\right\rangle _{\nu }$ & $\left\langle Q_{0}^{2}\right\rangle $ & $%
\left\langle Q_{0}^{2}\right\rangle _{\pi }$ & $\left\langle
Q_{0}^{2}\right\rangle _{\nu }$ & $\left\langle Q_{0}^{2}\right\rangle $ & \\ 

\hline

0.00 & 0.0 & 0.0 & 0.0 & 0.0 & 0.0 & 0.0 & 0.1689 \\ 

0.05 & -0.0025 & 0.0039 & 0.0015 & -0.0008 & 0.0057 & 0.0048 & 0.1709 \\ 

0.20 & -0.0087 & 0.0169 & 0.0082 & 0.1067 & 0.2100 & 0.3168 & 0.1636 \\ 

0.40 & -0.0100 & 0.0442 & 0.0342 & 0.0099 & 0.0701 & 0.0800 & 0.1624 \\ 

0.60 & 0.0218 & 0.1261 & 0.1479 & 0.0483 & 0.1617 & 0.2100 & 0.1655 \\ 

0.70 & 0.0683 & 0.2193 & 0.2876 & 0.0892 & 0.2455 & 0.3347 & 0.1682 \\ 

0.80 & 0.1416 & 0.3594 & 0.5010 & 0.4521 & 0.8958 & 1.3479 & 0.1713 \\ 

0.85 & 0.2227 & 0.5053 & 0.7280 & 11.2116 & 18.4734 & 29.6850 & 0.1432 \\ 

0.90 & 11.63 & 20.0514 & 31.6814 & 15.0534 & 25.0116 & 40.0650 & 0.1218 \\ 

0.95 & 14.9910 & 26.1956 & 41.1866 & 17.5444 & 29.7372 & 47.2816 & 0.0935 \\ 

1.00 & 17.4655 & 29.9152 & 47.3807 & 19.2454 & 33.1840 & 52.4295 & 0.0807 \\ 

1.05 & 22.3626 & 34.0604 & 56.4230 & 20.4735 & 35.9085 & 56.3820 & 0.0831 \\ 

1.15 & 31.6509 & 38.4407 & 70.0915 & 22.9707 & 39.9774 & 62.9481 & 0.0638 \\ 

1.20 & 33.9053 & 39.6519 & 73.5572 & 24.8922 & 41.7666 & 66.6589 & 0.0417 \\ 

\end{tabular}
\end{ruledtabular}
\end{table}

\end{document}